\documentclass[letterpaper,english,aps,prl,reprint,superscriptaddress]{revtex4-2}
\usepackage[T1]{fontenc}
\usepackage[latin9]{inputenc}
\usepackage{xcolor}
\usepackage{mathrsfs}
\usepackage{amsmath}
\usepackage{amssymb}
\usepackage{makeidx}
\makeindex
\usepackage{graphicx}

\makeatletter

\usepackage{hyperref}
\hypersetup{
    colorlinks = true,
	allcolors = blue	
}
\usepackage{times}

\makeatother

\usepackage{babel}
\begin{document}
\bibliographystyle{apsrev4-1.bst}

\title{Classification of Chern Numbers Based on High-Symmetry Points}

\author{Yu-Hao Wan}
\affiliation{International Center for Quantum Materials, School of Physics,
Peking University, Beijing 100871, China}
\author{Peng-Yi Liu}
\affiliation{International Center for Quantum Materials, School of Physics,
Peking University, Beijing 100871, China}
\author{Qing-Feng Sun}
\thanks{Corresponding author: sunqf@pku.edu.cn.}
\affiliation{International Center for Quantum Materials, School of Physics,
Peking University, Beijing 100871, China}
\affiliation{Hefei National Laboratory, Hefei 230088, China}

\begin{abstract}
The Chern number is a crucial topological invariant for distinguishing
the phases of Chern insulators.
Here we find that for Chern insulators with inversion symmetry,
the Chern number alone is insufficient to fully characterize their topology.
Specifically, distinct topological phases can be differentiated
based on skyrmions at different high-symmetry points.
Interfaces between these topological phases exhibit gapless helical states,
which provide counter-propagating transport channels and robust
quantized transport. Additionally, we identify topological transitions
that do not involve changes in the Chern number but can be characterized
by transitions of skyrmions between high-symmetry points. These transitions
arise due to the toroidal structure of the two-dimensional Brillouin
zone, which is generally applicable to two-dimensional periodic lattice system. Our research introduces new degrees of freedom for controlling topological optical transport and deepens the understanding of Chern insulators with inversion symmetry.
\end{abstract}

\maketitle
Over the past decade, topological insulators have been widely studied in photonic systems due to the robustness of their edge states against disorder\citep{haldane2008possible,mousavi2015topologically,sridhar2024quantized,wang2008reflectionfree,wu2015scheme,jiang2015signature}. Various topological states, such as the integer quantum Hall effect\citep{wang2008reflectionfree,wang2009observation,skirlo2015experimental,PhysRevLett.123.043201}, valley Hall effect\citep{dong2017valley,lu2018valley,gao2018topologically,noh2018observation,yang2021optically}, and spin Hall effect\citep{yin2013photonic,hafezi2011robust,bliokh2015quantum,leykam2018reconfigurable}, have been realized in different photonic crystal systems. The topological classification theory plays a pivotal role in this investigation by characterizing the topological phases of a Hamiltonian through distinct topological indices\citep{chiu2016classification}. For example, in photonic Chern insulators\citep{skirlo2015experimental,rechtsman2013photonic,fang2012realizing,wang2009observation}, the Chern number serves as a topological index that distinguishes different phases\citep{bansil2016emphcolloquium,aidelsburger2018artificial,wu2024edgea}.
Specifically, the absolute value of the Chern number reflects the
number of topological chiral boundary states, while its sign indicates
the direction of the edge flows, thereby defining
the chirality of edge states. 

\begin{figure}
\begin{centering}
\includegraphics[scale=0.4]{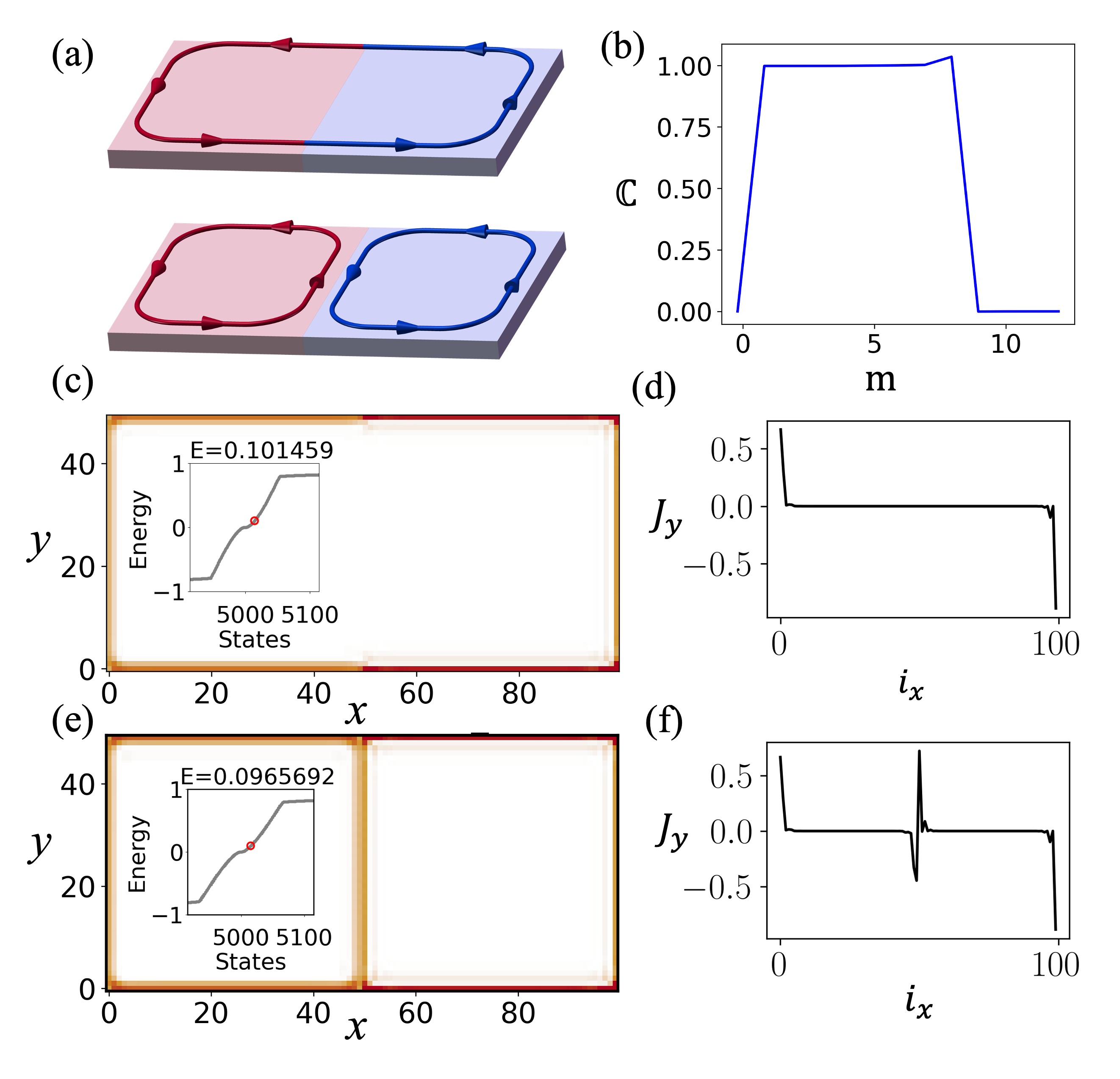}
\par\end{centering}
\caption{\protect\label{fig:1} (a) Schematic diagram: The red and blue regions
are set with different parameters, but both sides have a Chern number
of 1. Chiral edge states represented by red and blue lines. Top Image:
The edge states cancel each other out at the interface.
Bottom Image: The helical edge states exist at the interface.
(b) Chern number $\mathbb{C}$ as a function of the mass term $m$.
(c) Wave function distribution of the in-gap state (indicated by the red circle in the
inset), with $m_{l}=1$ and $m_{r}=2$. Inset: Energy versus the state index.
(d) Equilibrium current $J_{y}$ at energy $E=0.1$ as a function of $i_x$ in real space.
(e-f) Similar to (c-d), but with $m_{l}=1$ and $m_{r}=6$.}
\end{figure}

A key result of topological classification is the inevitable appearance of
gapless states at the interface between topologically distinct systems,
where the topological invariant changes. These gapless states manifest
as chiral modes\citep{rosen2017chiraltransport,zhao2023creation,zhang2024manipulation}.
The number of chiral modes at the interface of two systems with Chern
numbers $\ensuremath{\mathbb{C}_{1}}$ and $\mathbb{C}_{2}$ is determined by their
difference, denoted as $N_{\text{chiral}}=\mathbb{C}_{1}-\mathbb{C}_{2}$\citep{hasan2010colloquium,qi2011topological,zhao2023creation,rosen2017chiraltransport,zhang2024manipulation,wang2014universal,liu2016largediscrete,yan2024rulesfor}.
Specifically, when $\mathbb{C}_{1}=\mathbb{C}_{2}$, no gapless states appear at the
interface (top image in Fig.\ref{fig:1}(a)).

In this Letter, we discover that even for two systems with
the same Chern number, gapless helical states can still appear at the interface
(bottom image in Fig.\ref{fig:1}(a)), suggesting that these two systems
belong to different topological classes and the Chern number alone
cannot distinguish between them.
We point out that in systems with inversion symmetry,
those sharing the same Chern number can be further classified based on the high-symmetry points where the skyrmions are located.
This novel classification scheme enhances our understanding of the topological phases of Chern insulators.

We start from a 2-bands square lattice model with inversion symmetry,
\begin{align}
H= & \sum_{i_{x}}\sum_{i_{y}}[c_{\bf i}^{\dagger}\epsilon c_{\bf i}
+c_{\bf i}^{\dagger}T_{x}c_{{\bf i}+\delta x}
 +c_{\bf i}^{\dagger}T_{y}c_{{\bf i}+\delta y} \nonumber \\
 &+c_{\bf i}^{\dagger}T_{sec}c_{{\bf i}+\delta x +\delta y}
  +c_{\bf i}^{\dagger}T_{sec}c_{{\bf i}+\delta x -\delta y}+H.c.]\label{eq:1}
\end{align}
where $c_{{\bf i}}$ and $c_{{\bf i}}^{\dagger}$ are the
annihilation and creation operators at site ${\bf i}=(i_{x},i_{y})$, respectively.
Here $\delta x =(1,0)$, $\delta y =(0,1)$,
$\epsilon=(m-4B)\sigma_{z}$, nearest-neighbor hopping terms
$T_{x}=B\sigma_{z}$ and $T_{y}=\frac{A}{2i}\sigma_{y}+B\sigma_{z}$,
and the next-nearest-neighbor hopping term $T_{sec}=\frac{A}{4i}\sigma_{x}$.
$\hat{\sigma} =(\sigma_x, \sigma_y, \sigma_z)$ represents the Pauli matrices, $m$ is the Dirac mass term, and the parameters \( A \) and \( B \) are both set to 1 for convenience.
In momentum space, the Hamiltonian reduces as
\begin{equation}
H(\mathbf{k})=\mathbf{d}(\mathbf{k})\cdot\hat{\sigma}\label{eq:2}
\end{equation}
with $\mathbf{d}(\mathbf{k})=(A\sin k_{x}\cos k_{y},A\sin k_{y},m+2B(\cos k_{x}+\cos k_{y}-2))$ and $\mathbf{k}=(k_x,k_y)$.
This model is similar to the Qi-Wu-Zhang model,
which is the minimal model describing the Chern insulator\citep{qi2006topological}.
Both models have the same low-energy Dirac equation at the
$\Gamma$ point\citep{shen2012topological,shen2011topological} [see
Supplemental materials (SM)\citep{seesupplemental}].
The Chern number $\mathbb{C}$ can be calculated by integrating the Berry curvature
over the Brillouin zone. Figure \ref{fig:1}(b) shows the relationship
between $\mathbb{C}$ and the mass term $m$. The Chern number
remains 1 within the range of $m$ from $0$ to $8$,
but are they all the same topological phase?

To answer this question, we consider a rectangular system
with dimensions $L_{x}=100$ and $L_{y}=50$, where the mass term
differs in the left and right halves, as shown Fig.\ref{fig:1}(a).
Specifically, $m(i_{x})=m_{l}$ for $1\leq i_{x}\leq50$ and
$m(i_{x})=m_{r}$ for $50<i_{x} \leq 100$.
Therefore, the system can be viewed as two blocks
with different mass terms. Here, we fix $m_{l}=1$. For $m_{r}$,
we consider two cases: $m_{r}=2$ and $m_{r}=6$. The Chern number
is 1 for these three masses {[}See Fig. \ref{fig:1}(b){]}.

For the first case, where $m_{l}=1$ and $m_{r}=2$. By diagonalizing
the Hamiltonian, we obtain the energy levels of the system and
the wave function distribution of in-gap states in real space (see Fig. \ref{fig:1}(c)),
showing absence of modes at the interface between the left and right
blocks, which indicates that the two blocks are
in the same topological phase.
Figure. \ref{fig:1}(d) illustrates the distribution of the
equilibrium current $\ensuremath{J_{y}}(i_x)$ (See SM for detailed calculation
methods \citep{seesupplemental}).
We observe that equilibrium currents almost are zero
for $10<i_x<90$, reflecting no states at the interface.

For the second case where $m_{l}=1$ and $m_{r}=6$.
Energy levels and the wave function distribution
of the in-gap states are shown in Fig. \ref{fig:1}(e).
Surprisingly, localized modes appear at the interface between the two blocks.
Additionally, the calculation of $J_{y}(i_x)$ shows counter-propagating currents
at the interface [Fig. \ref{fig:1}(f)],
revealing the existence of the helical edge states.
The gapless helical states at the interface between the
two systems implies that they belong to different topological classes,
despite sharing the same Chern number. In this case, the Chern number
is insufficient to effectively distinguish the topology of the system.
\begin{figure}
  \begin{centering}
  \includegraphics[scale=0.45]{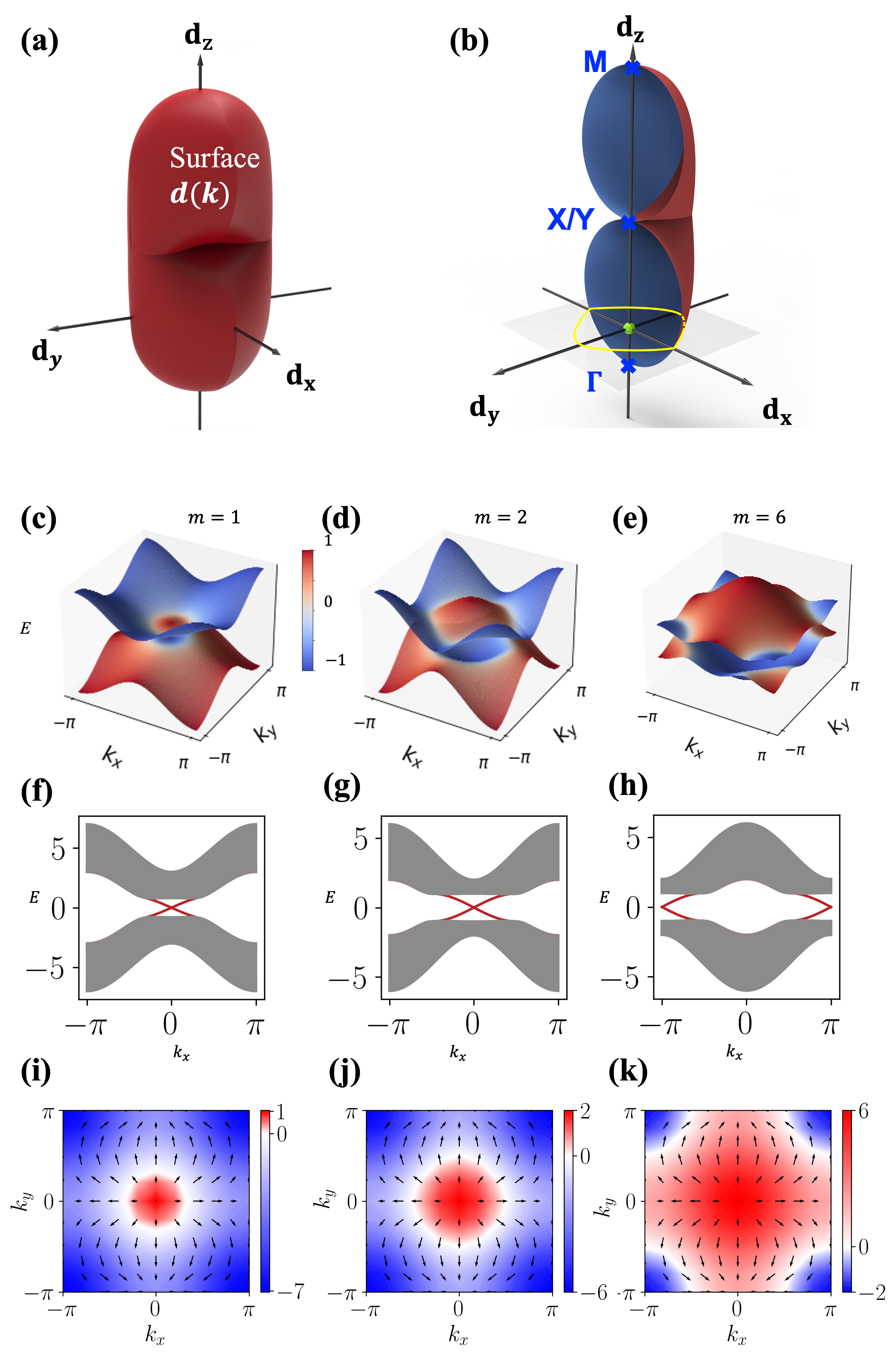}
  \par\end{centering}
  \caption{\protect\label{fig:2}
  (a)
  The surface $\ensuremath{\mathbf{d}(\mathbf{k})}$
  corresponding to Eq.(\ref{eq:2}) as ${\bf k}$ sweeps through the whole
  Brillouin zone.
  (b) The same surface $\mathbf{d}(\mathbf{k})$ as in
  (a), but only the image of half of the Brillouin zone is shown for
  clarity. The outer surface is red, the inner surface is blue, and
  the origin is represented by a green sphere. The yellow line indicates
  the BIS.
   (c-e) Bulk band structure for $m=1$, $2$,
  and $6$.
  (f-h) Band diagrams of quasi-one-dimensional nanoribbons corresponding to
  for $\ensuremath{m=1,2,6}$, with edge states marked in red.
  (i-k) The distribution of $\mathbf{d}(\mathbf{k})$ in the Brillouin zone for
  $\ensuremath{m=1,2,6}$. The vector field represents the components
  of $d_{x}$ and $d_{y}$, while color represents the $d_{z}$ component.}
  \end{figure}

  {
Why does Chern number classification fail here?}
To answer this question, let us review the definition of the Chern number. For a two-band
Bloch Hamiltonian $H(k)=\mathbf{d}(\mathbf{k})\cdot\hat{\sigma}$,
$\mathbf{d}(\mathbf{k})$ defines a vector field over the Brillouin
zone, and it also gives a mapping from the Brillouin zone $T_{2}$
to $S_{2}$ \citep{aidelsburger2018artificial}.
The vector field $\mathbf{d}(\mathbf{k})$ forms closed surfaces in three dimension,
which can wrap around the origin either positively or negatively.
In this picture, the Chern number quantifies how many times the surface
$\mathbf{d}(\mathbf{k})$ wrap around the origin\citep{asboth2016ashort}.

For example, the closed surfaces of $\mathbf{d}(k)$ corresponding
to Eq.(\ref{eq:2}) are illustrated in Fig. \ref{fig:2}(a).
For clarity, Fig. \ref{fig:2}(b) shows a sectioned view of the surface $\mathbf{d}(\mathbf{k})$ from (a), with the origin $(0,0,0)$ marked by a green sphere.
Note that the surface $\mathbf{d}(\mathbf{k})$ is divided into two
parts, each being a closed surface (see SM for details on $\mathbf{d}(k)$ \citep{seesupplemental}).
The mass term $m\sigma_{z}$ adjusts $d_{z}$, thereby shifting the
surface $\mathbf{d}(k)$ in the z-direction to control which closed part
wraps around the origin. When $0<m<4$ and $4<m<8$, the origin is
enclosed by the lower and upper parts, respectively.
However, in both cases, the surface $\mathbf{d}(k)$ wraps positively around
the origin once, resulting in a Chern number of 1 for both scenarios.
Nevertheless, these two cases should correspond to distinct topological phases
because the origin cannot be adiabatically moved from one closed region to
another.

The Chern number cannot distinguish between these two topological
phases because it is defined as the number of times the surface $\mathbf{d}(\mathbf{k})$ wraps around the origin\citep{asboth2016ashort}.
In this definition, the Chern number cannot differentiate which closed surface
surrounds the origin.
However, the origin being surrounded by which
closed surfaces determines in which topological phase the system resides.

{
Therefore, to distinguish the topologies here, we need more refined indicators to label which closed surface encloses the origin. In fact, for the system with inversion symmetry, this indicator corresponds to the high-symmetry points in the Brillouin zone.} {
To see this, we can examine the topological differences between the phases where \(0 < m < 4\) and \(4 < m < 8\).} Fig.\ref{fig:2}(c-e)
presents the bulk spectrum for $m=1$, $2$ and $6$, while the corresponding
quasi-one-dimensional nanoribbon system spectrum is shown in Fig.\ref{fig:2}(f-h).
In Fig.\ref{fig:2}(c-e), the color represents $|\psi_{I}|^{2}-|\psi_{II}|^{2}$,
with $\ensuremath{\psi_{I/II}}$ the two components of the spinor wave function.
This visualization clearly indicates the band
inversion surface (BIS) $d_{z}=0$ in white.
For $m=1$ and $2$, the BIS encircle the $\Gamma$ point,
corresponding to edge states around the $\Gamma$ point (red in Fig.\ref{fig:2}(f) and (g)).
For $m=6$, the BIS encircles the $M$ point, with corresponding edge states around
the $M$ point (Fig.\ref{fig:2}(h)).
In the cases of $m=1$ and $m=6$, the edge states have different momenta,
resulting in the existence of helical edge states at their interface in Fig. \ref{fig:1}(e) and (f).

To better visualize their topological differences, Fig.\ref{fig:2}(i-k)
shows the distribution of $\mathbf{d}(\mathbf{k})$ in the Brillouin
zone for $m=1$, $2$, and $6$. From the distribution of $\mathbf{d}(\mathbf{k})$,
we can identify skyrmion structures.
Skyrmion, a concept from field theory\citep{skyrme1962aunified},
play an important role in magnetism\citep{bogdanov1994thermodynamically,rossler2006spontaneous,muhlbauer2009skyrmion,yu2010realspace}
and topological classification due to its topological propertie\citep{qi2006topological,bjornson2014skyrmion,loder2017momentumspace}.
Previous studies have suggested a correlation between the number of
skyrmions and the Chern number\citep{bjornson2014skyrmion}. We can
determine a skyrmion's position using BIS and topological charges.
The topological charges, i.e., $d_{x}=d_{y}=0$, appear at the $\Gamma$,
$M$, $X$, and $Y$ points. These are a result of the system's inversion
symmetry, as will be proved later. For $m=1$ and $2$, the BIS encircles
the topological charge at the $\Gamma$ point, corresponding to a
skyrmion located at $\Gamma$. Conversely, for $m=6$ {[}Fig.\ref{fig:2}(k){]},
the BIS encircles the topological charge at the $M$ point, indicative
of a skyrmion located at $M$.

\begin{figure}
\begin{centering}
\includegraphics[scale=0.5]{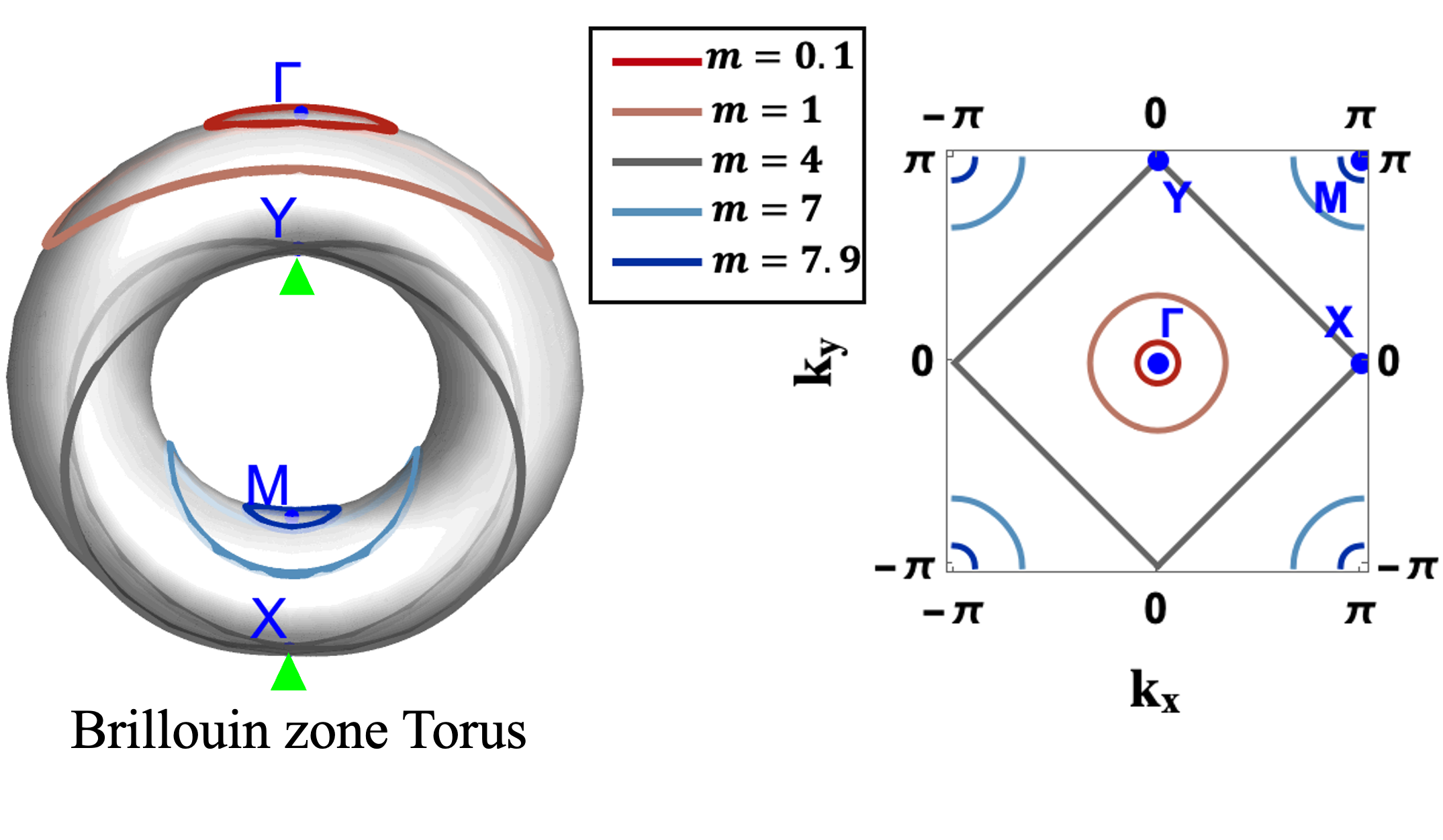}
\par\end{centering}
\caption{\protect\label{fig:3} Distribution of the BIS for different mass terms $m$ on
the 2D Brillouin zone torus (left) and the Brillouin zone plane (right).}
\end{figure}

In fact, skyrmions at different high-symmetry points can further classify the topology, as the origin being enclosed by different closed surfaces corresponds to skyrmions appearing at these distinct high-symmetry points.
To see this, we can examine how the skyrmion manifests on the surface $\mathbf{d}(\mathbf{k})$.
Similar to skyrmions in the Brillouin zone, skyrmions can
be identified by the BIS encircles a topological charge.
Here, the topological charge, i.e., $d_{x}=d_{y}=0$, corresponds to the intersection
of the surface with the $z$-axis (marked in blue crosses in Fig.\ref{fig:2}(b)).
The BIS is the plane $d_{z}=0$ intersecting with the surface $\mathbf{d}$(\textbf{k})
{[}yellow loop in the Fig.\ref{fig:2}(b){]}. A structure where a
topological charge is encircled by the BIS defines a complete skyrmion.
For instance, in Fig. \ref{fig:2}(b), the BIS encircles a charge
located at the $\Gamma$ point, indicating the presence of a skyrmion
at the $\Gamma$ point.

In our model, all charges appear at high-symmetry points. Below, we
demonstrate that this is due to inversion symmetry. The two-dimensional
inversion symmetry is written as $P=\hat{P}\otimes\hat{R}_{2D}$,
where $\hat{P}=\sigma_{z}$ acts on the spin space and $\hat{R}_{2D}$
is a 2D spatial operator acting on the real space $\mathbf{R}\rightarrow-\mathbf{R}$.
For the system with preserved inversion symmetry: $\hat{P}H(k)\hat{P}^{-1}=H(-k)$.
At the four high-symmetry points $\Lambda_{i}$, which are inversion-invariant
points, we have $\hat{P}H(\Lambda_{i})\hat{P}^{-1}=H(\Lambda_{i})$,
where $\{\Lambda_{i}\}=\{\Gamma(0,0),M(\pi,\pi),X(\pi,0),Y(0,\pi)\}$.
Since $H(\Lambda_{i})=\mathbf{d}(\Lambda_{i})\cdot\hat{\sigma}$,
we get $-d_{x/y}(\Lambda_{i})=d_{x/y}(\Lambda_{i})=0$, indicating
that topological charges must exist at these high-symmetry points.
We define the Chern number contributed by the skyrmion at the high-symmetry
point $\ensuremath{\Lambda_{i}}$ as $\ensuremath{\mathbb{C}_{\Lambda_{i}}}$.
Therefore, in the intervals $0<m<4$ and $4<m<8$, skyrmions appear
at the $\Gamma$ point and the $M$ point, respectively, corresponding
to $\mathbb{C}_{\Gamma}=1$ and $\mathbb{C}_{M}=1$.

In the surface $\mathbf{d}(\mathbf{k})$, lattice momentum $\ensuremath{{\bf k}}$
is a latent variable. To better illustrate the topological classification
based on high symmetry points, we directly observe the phase transitions
from the Brillouin zone, which explicitly contains momentum information.
Figure \ref{fig:3} shows the evolution of the BIS
in the Brillouin zone as a torus and as a plane when $\ensuremath{m}$
changes from 0.1 to 7.9. It can be observed that at $m=0.1$ and $m=7.9$,
the loops encircle the $\Gamma$ point (red circle) and the $M$ point
(blue circle), respectively, corresponding to skyrmions located at
the $\Gamma$ point and the $M$ point, defined as $\mathbb{C}_{\Gamma}=1$
and $\mathbb{C}_{M}=1$.

To transform a loop on the torus that encircles the $\Gamma$ point
into one that encircles the $M$ point, a straightforward method is
to translate the loop on the surface. However, this will inevitably
cause the loop to touch the $\Gamma$ point, which physically corresponds
to the gap closing and reopening at the $\Gamma$ point. Additionally,
when translating the loop, there is no guarantee that the $\Gamma$
point remains the inversion center of the loop, thus breaking the
inversion symmetry of the system.

However, the periodicity of the 2D Brillouin zone gives it a toroidal
geometric structure, leading to an alternative method to transform
the loop from encircling the $\Gamma$ point to encircling the $M$
point, without gap close at $\Gamma$ point and preserves inversion
symmetry throughout the process. By increasing the radius of the loop
(corresponding to increasing the mass term $m$ in the model), the
loop can grow until it touches itself on the opposite side of the
torus (marked in green triangles). This discontinuous change corresponds
to a topological transition, and after this transition, the loop becomes
one that encircles the $M$ point. The phase transition here relies
on the toroidal structure of the Brillouin zone. Considering that any two-dimensional periodic lattice system features a Brillouin zone with a toroidal structure, such phase transitions are likely to be universally present.

\begin{figure}
\begin{centering}
\includegraphics[scale=0.5]{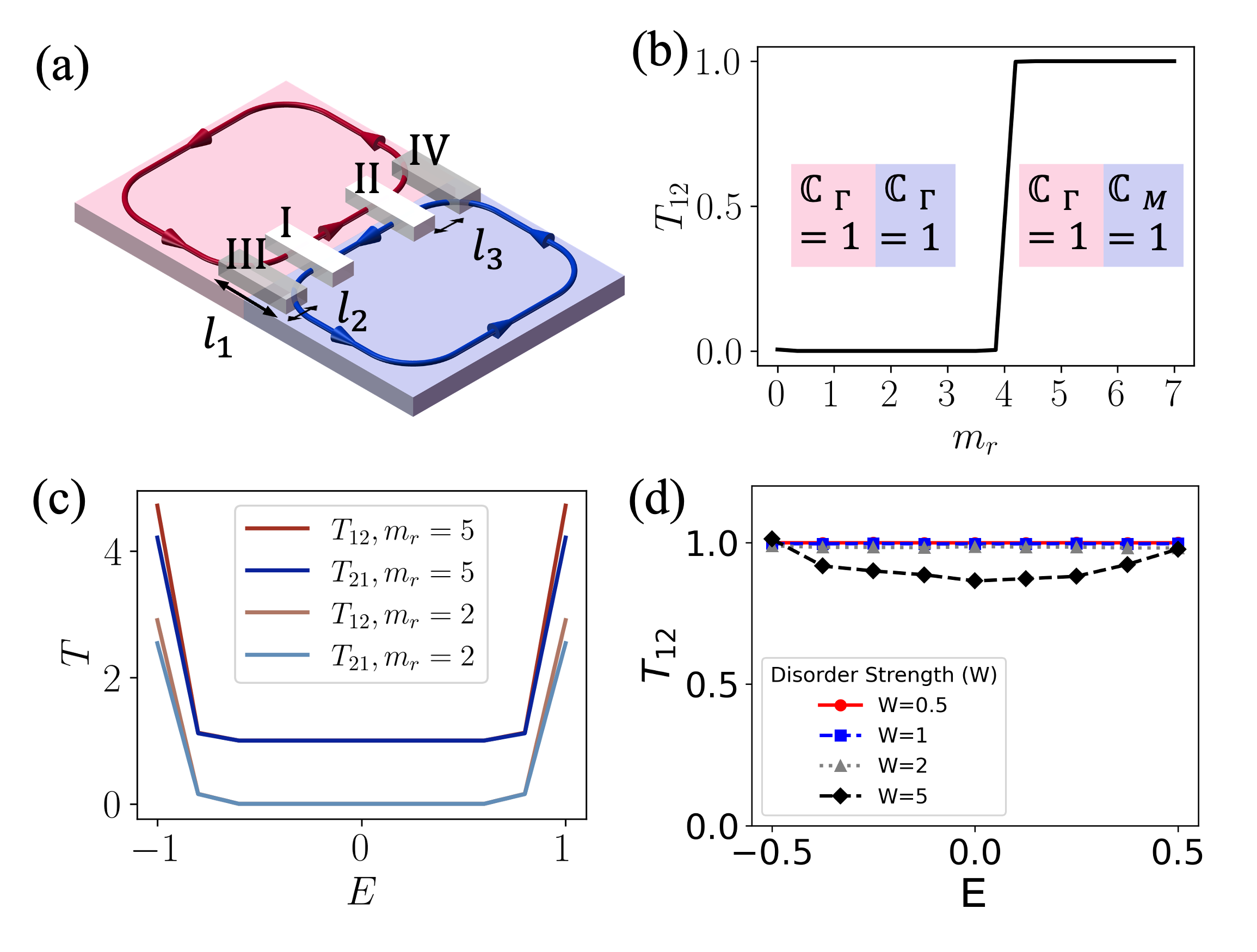}
\par\end{centering}
\caption{\protect\label{fig:4} (a) Schematic of the transport device:
Measurement ports are labeled I and II, while protective ports are labeled III
and IV. The geometric parameters are $l_{1}=10$, $l_{2}=5$, and $l_{3}=10$.
(b) Dependence of $T_{12}$ on $m_{r}$ at energy $E=0$.
The inset marks the Chern numbers $\mathbb{C}$ base on high-symmwtry points
of the left and right halves when \( m_{r} \) is in the intervals 0 to 4 and 4 to 7, respectively.
(c) Plots of $T_{12}$ and $T_{21}$ versus energy for different values of $m_{r}$.
(d) $T_{12}$ versus energy for different disorder strengths $W$ with $m_{r}=5$}
\end{figure}

The helical modes at the interface can be detected through
transport measurements. We constructed the rectangular system shown in
Fig.\ref{fig:4}(a), with dimensions $L_{x}=100$ and $L_{y}=50$.
The system is divided into left and right parts at $i_{x}=50$, with
different mass term. The mass term for the left part are set to $m_{l}=1$,
while the mass term for the right part set to $m_{r}$. The prots
are attached on the interface of the two blocks, labeled as I-IV.
Ports III and IV are grounded to prevent the influence of the outer chiral
channel. The transmission probability between ports I and II reflects
the number of channels at the interface.

The transmission $T$ between port $n$ and port $m$ can be derived
by using non-equilibrium Green's function\citep{wang2016landauer,salerno2019quantized,buttiker1988absence,fisher1981relation,landauer1970electrical,gong2020transport}:
$T_{mn}(E)=\operatorname{Tr}\left[\Gamma_{n}\mathbf{G}^{R}\Gamma_{m}\mathbf{G}^{A}\right]$
where $\Gamma_{n/m}$ represents the line-width function of the $n/m$-th
port, which is a constant under the wide-band approximation. The retarded
Green's function is defined as: $\mathbf{G}^{R}(E)=\left[\mathbf{G}^{A}\right]^{\dagger}=\left[(E+{i}0^+)\mathbf{I}-\mathbf{H}-\sum_{n}\Sigma_{n}^{R}\right]^{-1}$.
The self-energy term is given by: $\Sigma_{n}^{R}=-\frac{{i}}{2}\Gamma_{n}$.

Figure \ref{fig:4}(b) shows the variation of $T_{12}$ as $m_{r}$
increases. At $m_{r}=4$, $T_{12}$ suddenly changes from 0 to
1. This indicates that around $m=4$, as the right side transitions
from the $\mathbb{C}_{\Gamma}=1$ phase to the $\mathbb{C}_{M}=1$ phase, gapless helical modes
appear at the interface
because the right and left sides belong to different topological phases.
Figure \ref{fig:4}(c) shows the transmission
probabilities $T_{12}$ and $T_{21}$ as a function of the energy
for different $m_{r}$. When $m_{r}=2$, both sides of the system
belong to $\mathbb{C}_{\Gamma}=1$, and no gapless states are present at the
interface, resulting in a transmission probability of 0 within the
gap. However, when $m_{r}=5$,
the left and right sides are at the different topological phases,
leading to the presence of the helical transport channel
at the interface and $T_{12}=T_{21}=1$.

To investigate the robustness of the helical modes at
the interface, the on-site energies at the Hamiltonian in Eq.(\ref{eq:1}) are added with a disorder term $\epsilon_{i}\sigma_{0}$
with\citep{kawarabayashi2009quantum,cheng2016thevalley,cheng2011effectof,yang2018gatevoltage}
\[
\varepsilon_{\bf i}=\left.{\sum_{\bf j}\tilde{\varepsilon}_{\bf j}
e^{-\left|\mathbf{r}_{\bf ij}\right|^{2}/2\eta^{2}}}\right/{\sqrt{\sum_{\bf j}
e^{-\left|\mathbf{r}_{\bf ij}\right|^{2}/2\eta^{2}}
}},
\]
Here $\tilde{\varepsilon}_{j}$ is uniformly distributed in the interval $[-W/2,W/2]$,
where $W$ the disorder strength.
$\left|\mathbf{r}_{\bf ij}\right|$ denotes
the distance between site ${\bf i}$ and ${\bf j}$, while $\eta$ describes the
correlation length of the disorder. In the numerical calculations,
we use a long-range disorder with $\eta=1.5a$, where $a$ is the
lattice constant, set to 1 in this context. The transmission coefficient
$T_{12}(E)$ is averaged over 100 different disorder configurations
for each value of $W$. Figure \ref{fig:4}(d) illustrates
$T_{12}(E)$ under varying $W$. It remains quantized
even when $W=1$, indicating effective suppression of backscattering.
This demonstrates the robustness of the topologically protected counter-propagating helical modes at the interface, supported respectively by $\mathbb{C}_{\Gamma}=1$
and $\mathbb{C}_{M}=1$.

In addition to photonic systems, our classification theory is broadly applicable to all inversion-symmetric systems with a Brillouin zone, including cold atoms, acoustic, and mechanical systems. In fact, the Qi-Wu-Zhang model, which closely resembles the model described by Eq.(\ref{eq:2}) in this study, has already been realized in Bose-Einstein condensates \citep{wang2018diracrashba,sun2018highlycontrollable,sun2018uncover,wu2016realization} and optical Raman lattices \citep{liang2023realization}. Additionally, BIS, which determines the position of skyrmions, can be both observed and adjusted experimentally \citep{sun2018uncover}, and recent advancements in real-space detection based on ultracold atoms have enabled the experimental observation of topological modes at the interface between two distinct topological systems \citep{braun2024realspace}.

In summary, we propose that for systems with inversion symmetry, those
with the same Chern number can be further classified based on the
high-symmetry points where the skyrmions are located.
On the torus surface of the Brillouin zone,
the topological transition between different phases with the same Chern number
involves that the loop of the BIS touches itself and reopens.
Gapless helical states appear at the interface between different
topological phases, providing counter-propagating transport channels. These transport channels are topologically protected and
robust against disorder, offering additional avenues for controlling topological optical transport. Furthermore, our work has deepened the understanding of the topological phases of Chern insulators with inversion symmetry.

\begin{acknowledgments}
  Y.-H. W. is grateful to Jiayu
Li, Ming Gong, Fajie Wang, Yucheng Wang, Chao Yang and Ludan Zhang for fruitful discussions.
This work was financially supported by
the National Natural Science Foundation of China (Grant No. 12374034 and No. 11921005),  the Innovation Program for Quantum Science and Technology
(2021ZD0302403), and the Strategic priority Research
Program of Chinese Academy of Sciences (Grant No.
XDB28000000). We also acknowledge the Highperformance
Computing Platform of Peking University for
providing computational resources.
\end{acknowledgments}
\bibliography{ref}

\end{document}